\documentclass[runningheads]{llncs}
\usepackage[T1]{fontenc}
\usepackage{graphicx}
\usepackage[misc]{ifsym}
\usepackage[linesnumbered,ruled]{algorithm2e}
\usepackage{booktabs}
\usepackage{multirow}
\usepackage{float}
\usepackage{subfig}
\usepackage{overpic}
\usepackage{bbding}
\begin{document}
\title{PadAug: Robust Speaker Verification with Simple Waveform-Level  Silence Padding}
%
%
\author{Zijun Huang\inst{1,3} \and
Chengdong Liang\inst{2} \and
Jiadi Yao\inst{1,3} \and
Xiao-Lei Zhang\inst{1}\textsuperscript{(\Letter)}\inst{,3}}
\authorrunning{Z. Huang et al.}
%
\institute{School of Marine Science and Technology, Northwestern Polytechnical University, Xi'an, China\\
\email{xiaolei.zhang@nwpu.edu.cn}\and
Guasemi, Shanghai, China
\and
Shenzhen Research Institute of Northwestern Polytechnical University,\\ Shenzhen, China 
}
\maketitle              
\begin{abstract}
The presence of non-speech segments in utterances often leads to the performance degradation of speaker verification. Existing systems usually use voice activation detection as a preprocessing step to cut off long silence segments. However, short silence segments, particularly those between speech segments, still remain a problem for speaker verification. To address this issue, in this paper, we propose a simple wave-level data augmentation method, \textit{PadAug}, which aims to enhance the system's robustness to silence segments. The core idea of \textit{PadAug} is to concatenate silence segments with speech segments at the waveform level for model training. Due to its simplicity, it can be directly applied to the current state-of-the art architectures. Experimental results demonstrate the effectiveness of the proposed \textit{PadAug}. For example, applying \textit{PadAug} to ResNet34 achieves a relative equal error rate reduction of 5.0\% on the voxceleb dataset. Moreover, the \textit{PadAug} based systems are robust to different lengths and proportions of silence segments in the test data.

\end{abstract}
\section{Introduction}
Speaker Verification (SV) aims at determining whether a speech utterance is uttered from an enrolled speaker or not. A conversational SV system mainly consist of two parts:  an embedding vector extractor \cite{dehak2010front} which produces discriminative embeddings for utterances of different speakers, and a back-end classifier, e.g. PLDA \cite{ioffe2006probabilistic} or Cosine similarity scoring, which computes the similarity score between the testing utterances and erollment utterances. Currently, deep-based SV systems \cite{bai2021speaker,li2017deep,snyder2018x} have achieved the state-of-the-art performance in many scenarios such as intelligent housing systems \cite{yao2023branch}, adversarial attack \cite{yao2023symmetric}, anti-spoofing \cite{chen2022masking,zhangdoubledeceiver}. However, their performance were still affected strongly by silence segments \cite{song2023trimtail,hasan2013duration}, since that the acoustic features of silence segments behave like additive noise of speech when generating speaker embeddings via e.g. x-vectors.

A common way of dealing with the silence segments is to use a voice activation detection (VAD) \cite{sohn1999statistical} based front-end to filter out silence segments. VAD usually uses a statistical model to model the characteristics of speech and non-speech signals such as energy, and then uses a frame dropping strategy to detect long silence segments. However, to avoid dropping unvoiced speech e.g. consonants, VAD always uses hang-before and hangover criteria \cite{vlaj2009influence} to keep short periods around voiced speech, where the short periods may be silence segments that affect the robustness of SV. 
 
 To deal with the short silence periods, another common approach is the attentive pooling, e.g. \cite{desplanques2020ecapa}, which allocates high weights to the frames that are important to SV embeddings, behaves like an implicit VAD in SV. However, attentive pooling considers too much information for generating the SV embeddings, instead of merely removing the negative effect of the short silence periods.

In this paper, we resort another simple way to address the above issue---data augmentation. 
This research direction seems far from explored yet. To our knowledge, existing studies mostly focus on improving the robustness of speech processing systems against noise environments only. For example, \cite{seltzer2013investigation} synthesised noisy data by superimposing clean speech and noise for speech recognition. \cite{ko2015audio} introduced speed perturbation on raw speech for speech recognition. \cite{kim2017generation} explored simulated data with an acoustic room
simulator in far-field ASR systems. \cite{park2019specaugment} proposed spectrum augmentation to the log mel spectrogram of data. \cite{song2023trimtail} applied length penalty to improve the latency of streaming speech recognition models.
\cite{snyder2018x} proposed to add additive noise and reverberation to speech for improving the robustness of SV in various noisy conditions.
\cite{kim2023pas} proposed partial additive noise to enhance the robustness of SV towards background noise.

Given the aforementioned analysis, motivated by \cite{song2023trimtail}, this paper proposes a simple and effective data augmentation method, named silence padding augmentation (\textit{PadAug}). \textit{PadAug} can be simply described as that it directly concatenates silence segments with speech utterances for training. Essentially, it makes a SV model biased towards learning the discriminative information between speaker segments and non-speaker segments, thus avoiding putting too much focus on the silence segments.

Different from existing data augmentation methods of SV, which directly adds additive noise to speech, e.g. partial additive noise \textit{PAS} \cite{kim2023pas}, for enhancing the robustness of SV systems in noisy environment (see Section \ref{sec:related} for more related work), \textit{PadAug} aims at enhancing the performance of SV in scenarios where a test utterance contains silence segments that appear in random positions with different lengths.

We applied \textit{PadAug} on several popular end-to-end SV networks to verify the effectiveness of the proposed method on improving the robustness of SV against silence segments. Extensive experimental results on Voxceleb demonstrate the effectiveness of the proposed method.

\section{Related work}\label{sec:related}
\subsection{Waveform augmentation}

The data augmentation strategies in \cite{snyder2018x} were widely used to enhance the performance for noisy scenarios, where each utterance file is modified by the following approaches:

\textit{Additive noises}: Music, Babble, and Noise files from MUSAN dataset \cite{snyder2015musan} are randomly selected and added add at a SNR margin. \cite{kim2023pas} extended the method to partial additive noises by controlling the duration of noise segments to further enhance the robustness of SV in noisy environments.

\textit{Reverberation}: Simulated room impulse responses (RIRs) are used to add reverberation into the speech recordings \cite{ko2017study}. Many papers considered using the RIR\_NOISES dataset \cite{habets2006room} to generate the reverberant training data.

\subsection{Spectrogram augmentation}
 Data augmentation on spectrograms was proposed by \cite{park2019specaugment} for speech recognition. The spectrograms of speaker utterances are augmented by \textit{time warping}, \textit{time masking} and \textit{frequency masking}. \cite{song2023trimtail} applied \textit{length penalty} to the spectrograms by trimming the trailing frames, leading frames and corresponding padding frames of the spectrograms.

\section{Proposed method}
Motivated by the length penalty and attention mechanism, \textit{PadAug} aims to augment data at the waveform-levels, which is directly applied to various SV systems for enhancing the robustness of SV against silence segments.

\begin{figure}[!t]
        \centering
        \subfloat[\label{fig:nopad}]{
                \includegraphics[scale=0.37]{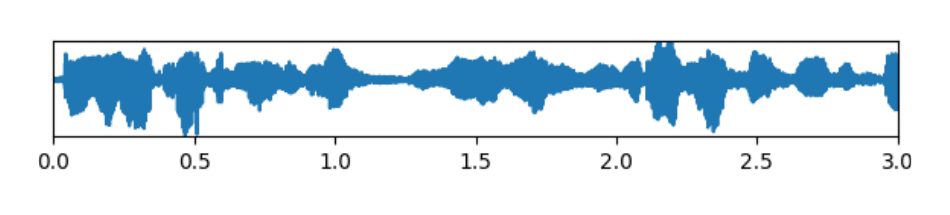}
        }
        \subfloat[\label{fig:frontpad}]{
                \includegraphics[scale=0.37]{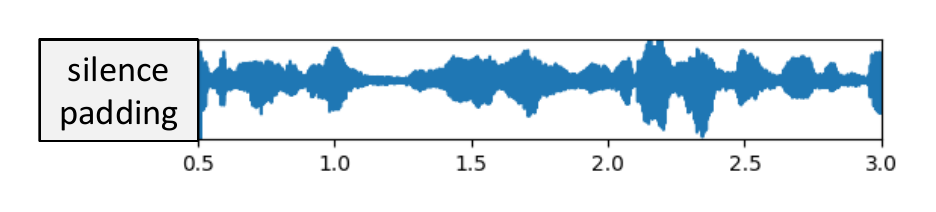}
        }
        \\
        \subfloat[\label{fig:endpad}]{
                \includegraphics[scale=0.37]{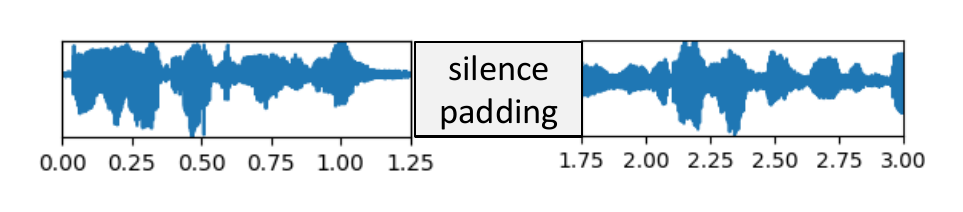}
        }
        \subfloat[\label{fig:midpad}]{
                \includegraphics[scale=0.37]{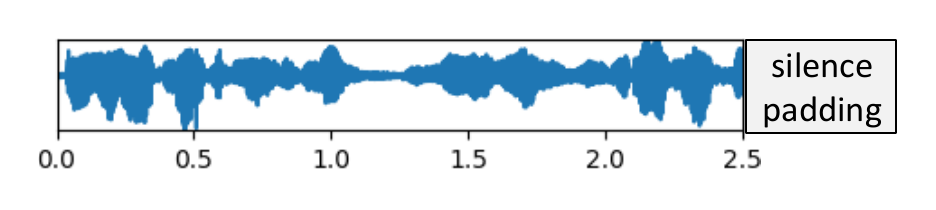}
        }
        \caption{ Illustration of PadAug: From top to bottom: (a) Waveform of the utterance without silence padding, (b) Waveform of the utterance with silence padding on the head, (c) Waveform of the utterance with silence padding on the central body, (d) Waveform of the utterance with silence padding on the tail. }
        \label{fig1}
\end{figure}

As shown in Figure \ref{fig1}, \textit{PadAug} modifies the input data of speaker by trimming the waveform and adding silence segments. The silence padding consists of Gaussian white noise and randomly concatenates the speaker segments on the head, central body and tail. Algorithm \ref{alg:PadAug} gives the details of the proposed data augmentation progress.

\begin{algorithm}[t]
\caption{Applying \textit{PadAug} to a mini-batch}\label{alg:PadAug}
    \KwIn{mini-batch $X$, noise $N$, minimum duration $T_{min}$, maximum duration $T_{max}$, $\mathrm{SNR}_{min}$, $\mathrm{SNR}_{max}$, $use\_mid$.}
    \KwOut{training mini-batch $X^{'}$}
    \tcc{data is temporal sequence}
$X^{'} = []$\;
\For{$x$ in $X$}{
    Sample $ T_{s} \sim \mathrm{Randint}(T_{min}, T_{max})$\;
    $x \gets \mathrm{Random}\_\mathrm{Chunk}(x,T_{s})$\;
    $L_{pad} \gets T_{max} - T_{s}$\;
    Sample $L_{head} \sim \mathrm{Randint}(0, L_{pad})$\;
    \eIf {$use\_mid$}
    {
    Sample $L_{mid} \sim \mathrm{Randint}(0, L_{pad}-L_{head})$\;
    $L_{tail} \gets L_{pad} - L_{head} - L_{mid}$\;
    }{
    $L_{mid} \gets 0$\;
    $L_{tail} \gets L_{pad} - L_{head}$\;
    }
    Sample $\mathrm{SNR} \sim \mathrm{Randint}(\mathrm{SNR}_{min},\mathrm{SNR}_{max})$\;
    $n \leftarrow \mathrm{wgn}($x$,\mathrm{SNR})$\;
    $\mathrm{Seg}_{head} \gets n(0,L_{head})$\;
    $\mathrm{Seg}_{mid} \gets n(L_{head}, L_{head}+L_{mid})$\;
    $\mathrm{Seg}_{tail} \gets n(L_{head}+L_{mid}, L_{pad})$\;
    Sample $P_{mid} \sim \mathrm{Randint}(0,T_{s})$\;
    $x_{left} \gets X(0,P_{mid})$\;
    $x_{right} \gets X(P_{mid},T_{s})$\;
    $x^{'} \gets \mathrm{concat}(\mathrm{Seg}_{head},x_{left},\mathrm{Seg}_{mid},x_{right},\mathrm{Seg}_{tail})$\;
    $X^{'} \gets \mathrm{append}(X^{'},x^{'})$\;
}
\end{algorithm}

\section{Experiments}
\subsection{Datasets}
We used the VoxCeleb dataset \cite{nagrani2017voxceleb,chung2018voxceleb2} for experiments. We trained the speaker verification networks on the development set of Vox-Celeb2 \cite{chung2018voxceleb2} which consists of 1,092,009 utterances from 5,994 speakers. We used three types of evaluation sets VoxCeleb1-O, VoxCeleb1-H and VoxCeleb1-H, which are drawn from the VoxCeleb1 training set \cite{nagrani2017voxceleb} for evaluation. We adopted the noise datasets from MUSAN \cite{snyder2015musan} and RIRs \cite{habets2006room} for the conventional data augmentation in the model training.

To construct the training data, each utterance of Voxceleb was clipped into a 3-second segment, where the voiced speech was guaranteed to be  from 1 second to 3 seconds. To simulate the real-world scenario where an utterance is short and filled with random silence segments, we constructed the following test datasets:
\begin{itemize}
\item Original: The original test sets, which are VoxCeleb1-O, VoxCeleb1-H and VoxCeleb1-H respectively.
    \item Chunk3s: The test utterances were clipped into segments of 3 seconds long.
    \item Chunk3s+Head1s+Tail1s: Each 3-second utterance in Chunk3s was padded with a 1-second silence at the head and tail respectively.
    \item Chunk3s+Head1s+Tail1s+Mid1s: Each 3-second utterance in Chunk3s was padded with a 1-second silence on the head, middle, and tail respectively.
\end{itemize}

\begin{table}[!t]
\caption{Comparison between the SV models with the proposed \textit{PadAug(HT)} and those without the proposed method.}
\centering
\label{tlb:1}
\scalebox{0.73}{
\begin{tabular}{lcccccccc}
\toprule
\multirow{2}{*}{SV Model}           & \multirow{2}{*}{Augmentation method} & \multirow{2}{*}{Test set} & \multicolumn{2}{c}{Voxceleb1-O} & \multicolumn{2}{c}{Voxceleb1-E} & \multicolumn{2}{c}{Voxceleb1-H} \\
\cmidrule(lr){4-5}\cmidrule(lr){6-7}\cmidrule(lr){8-9}
                                 &                          &                        & EER            & minDCF         & EER            & minDCF         & EER            & minDCF         \\
\midrule
\multirow{6}{*}{ResNet34} & \XSolidBrush                        & Original                      & 0.962& 0.091& 1.007& 0.110& 1.859& 0.178\\
                                 & PadAug(HT)                   & Original                      & 0.914& 0.077& 0.978& 0.104& 1.798& 0.174\\
                                 & \XSolidBrush                        & Chunk3s                & 1.345& 0.143& 1.259& 0.142& 2.348& 0.227\\
                                 & PadAug(HT)                   & Chunk3s                & 1.170& 0.114& 1.261& 0.140& 2.326& 0.231\\
                                 & \XSolidBrush                        & Chunk3s+Head1s+Tail1s  & 1.409& 0.142& 1.415& 0.156& 2.560& 0.249\\
                                 & PadAug(HT)                   & Chunk3s+Head1s+Tail1s  & 1.170& 0.122& 1.261& 0.144& 2.311& 0.229\\
\midrule
\midrule

\multirow{6}{*}{ECAPA-TDNN-512} & \XSolidBrush                        & Original                      & 1.090          & 0.115          & 1.227          & 0.134          & 2.278          & 0.219          \\
                                 & PadAug(HT)                   & Original                      & 1.064          & 0.098          & 1.174          & 0.129          & 2.177          & 0.207          \\
                                 & \XSolidBrush                        & Chunk3s                & 1.468          & 0.162          & 1.526          & 0.176          & 2.831          & 0.275          \\
                                 & PadAug(HT)                   & Chunk3s                & 1.526          & 0.180          & 1.557          & 0.172          & 2.828          & 0.276          \\
                                 & \XSolidBrush                        & Chunk3s+Head1s+Tail1s  & 1.781          & 0.208          & 1.864          & 0.205          & 3.399          & 0.319          \\
                                 & PadAug(HT)                   & Chunk3s+Head1s+Tail1s  & 1.435          & 0.162          & 1.569          & 0.174          & 2.831          & 0.278          \\
\midrule
\midrule
\multirow{6}{*}{ECAPA-TDNN-1024} & \XSolidBrush                        & Original                      & 0.909          & 0.083          & 1.056          & 0.113          & 1.942          & 0.189          \\
                                 & PadAug(HT)                   & Original                      & 0.830          & 0.088          & 0.995          & 0.108          & 1.845          & 0.183          \\
                                 & \XSolidBrush                        & Chunk3s                & 1.175          & 0.121          & 1.313          & 0.147          & 2.461          & 0.248          \\
                                 & PadAug(HT)                   & Chunk3s                & 1.165          & 0.135          & 1.316          & 0.150          & 2.447          & 0.242          \\
                                 & \XSolidBrush                        & Chunk3s+Head1s+Tail1s  & 1.473          & 0.159          & 1.547          & 0.175          & 2.899          & 0.280          \\
                                 & PadAug(HT)                   & Chunk3s+Head1s+Tail1s  & 1.154          & 0.144          & 1.327          & 0.152          & 2.411          & 0.241          \\
\midrule
\midrule
\multirow{6}{*}{CAM++}           & \XSolidBrush                        & Original                     & 0.723          & 0.122          & 0.940          & 0.109          & 1.916          & 0.194          \\
                                 & PadAug(HT)                   & Original                      & 0.707          & 0.097          & 0.911          & 0.106          & 1.818          & 0.178          \\
                                 & \XSolidBrush                        & Chunk3s                & 1.042          & 0.155          & 1.187          & 0.143          & 2.388          & 0.240          \\
                                 & PadAug(HT)                   & Chunk3s                & 0.963          & 0.147          & 1.156          & 0.136          & 2.302          & 0.238          \\
                                 & \XSolidBrush                        & Chunk3s+Head1s+Tail1s  & 1.249          & 0.184          & 1.387          & 0.171          & 2.767          & 0.281          \\
                                 & PadAug(HT)                   & Chunk3s+Head1s+Tail1s  & 1.000          & 0.150          & 1.172          & 0.138          & 2.281          & 0.232         \\
\toprule
\end{tabular}
}
\end{table}

\begin{table}[t]
\caption{Comparison between \textit{PadAug(HT) } and \textit{PadAug(HMT)}, where the SV model is ResNet34.}
\label{tlb:2}\centering
\scalebox{0.75}{
\begin{tabular}{lccccccc}
\toprule
\multirow{2}{*}{Method} & \multirow{2}{*}{Testset type}     & \multicolumn{2}{c}{Voxceleb1-O} & \multicolumn{2}{c}{Voxceleb1-E} & \multicolumn{2}{c}{Voxceleb1-H} \\
\cmidrule(lr){3-4}\cmidrule(lr){5-6}\cmidrule(lr){7-8}
                          &                             & EER            & minDCF         & EER            & minDCF         & EER            & minDCF         \\
\midrule
no \textit{PadAug}              & Original                           & 0.962          & 0.091          & 1.007          & 0.110          & 1.859          & 0.178          \\
PadAug(HT)   & Original                           & 0.914          & \textbf{0.077}          & \textbf{0.978}          & \textbf{0.104}          & \textbf{1.798}          & 0.174          \\
PadAug(HMT)         & Original                           & \textbf{0.904}          & \textbf{0.077}          & 0.990          & 0.107          & 1.809          & \textbf{0.173}          \\
\midrule
\midrule
no \textit{PadAug}              & Chunk3s                     & 1.345          & 0.143          & \textbf{1.259}          & 0.142          & 2.348          & \textbf{0.227}          \\
PadAug(HT)          & Chunk3s                     & \textbf{1.170}          & \textbf{0.114}          & 1.261          & \textbf{0.140}          & \textbf{2.326}          & 0.231          \\
PadAug(HMT)         & Chunk3s                     & 1.233          & 0.125          & 1.278          & 0.144          & 2.345          & 0.234          \\
\midrule
\midrule
no \textit{PadAug}              & Chunk3s+Head1s+Tail1s       & 1.409          & 0.142          & 1.415          & 0.156          & 2.560          & 0.249          \\
PadAug(HT)          & Chunk3s+Head1s+Tail1s       & \textbf{1.170}          & 0.122          & \textbf{1.261}          & 0.144          & \textbf{2.311}          & \textbf{0.229}          \\
PadAug(HMT)         & Chunk3s+Head1s+Tail1s       & 1.207          & \textbf{0.120}          & 1.269          & \textbf{0.142}          & 2.328          & 0.230          \\
\midrule
\midrule
no \textit{PadAug}              & Chunk3s+Head1s+Tail1s+Mid1s & 1.526          & 0.169          & 1.512          & 0.170          & 2.724          & 0.265          \\
PadAug(HT)          & Chunk3s+Head1s+Tail1s+Mid1s & \textbf{1.213}          & \textbf{0.133}          & 1.309          & 0.147          & 2.365          & \textbf{0.227}          \\
PadAug(HMT)          & Chunk3s+Head1s+Tail1s+Mid1s & 1.223          & 0.146          & \textbf{1.302}          & \textbf{0.144}          & \textbf{2.363}          & 0.234          \\
\toprule
\end{tabular}
}
\end{table}

\subsection{Comparison methods}

For the proposed \textit{PadAug}, our experiments contain two position strategies:
\begin{itemize}
    \item \textbf{Head-tail based \textit{PadAug} (PadAug(HT))}:  Silence paddings were concatenated at the head and tail of each speech segment.

    \item \textbf{Head-mid-tail based \textit{PadAug} (PadAug(HMT))}: Silence paddings were concatenated at the head, body and tail of each speech segment.
\end{itemize}
Note that the length of each padding component was random. We adopted the proposed augmentation method on four popular and state-of-the-art architectures: Resnet-34 \cite{zeinali2019but}, ECAPA-TDNN-512 \cite{snyder2019speaker}, ECAPA-TDNN-1024 \cite{garcia2020jhu}, and CAM++ \cite{wang2023cam++}. We implemented the models by ourselves to maintain a fair comparison on the investigation of whether using the proposed method or not.

We took the no-padding strategy in Figure \ref{fig1} (a) as the most important baseline, which only adds additive noise or reverberation to the input data with a probability of 0.6. 






\setcounter{footnote}{0} 

\subsection{Implementation details}
We used the open-source speaker embedding learning toolkit Wespeaker \footnote{https://github.com/wenet-e2e/wespeaker}\cite{wang2023wespeaker} to implement the proposed data augmentation method and baselines. Speed perturbation was conducted by randomly changing the speed of the utterances with a ratio of 0.9 or 1.1. We adopted 80-dimensional log Mel-filter banks (Fbank) which were pre-emphasized by a Hamming window with a width 25ms and a window shift of 10ms as the input features. All training segments were chucked into 300 frames and each was normalized by the cepstral mean normalization (CMN). The loss functions of all models were the angular additive margin softmax (AAM-Softmax) \cite{deng2019arcface}, where the scale was set to 32, and the initial margin was set to 0 and updated until 0.2 by the margin schedule \cite{wang2023wespeaker}.  The initial learning rate was set to $10^{-1}$. We used the warm-up and exponential decrease schedule in \cite{wang2023wespeaker} to update the learning rate until $5\times 10^{-5}$.

\begin{table}[!t]
\caption{Comparison between VAD and the proposed \textit{PadAug(HT)} on the 'chunk3s+head1s+tail1s' test sets.}
\centering
\label{tlb:3}
\scalebox{0.8}{
\begin{tabular}{lccccccc}
\toprule
\multirow{2}{*}{Model}           & \multirow{2}{*}{Approch} & \multicolumn{2}{c}{Voxceleb1-O} & \multicolumn{2}{c}{Voxceleb1-E} & \multicolumn{2}{c}{Voxceleb1-H} \\
\cmidrule(lr){3-4}\cmidrule(lr){5-6}\cmidrule(lr){7-8}
                                 &                                                  & EER            & minDCF         & EER            & minDCF         & EER            & minDCF         \\
\midrule
\multirow{2}{*}{ResNet34}
                                 & VAD                        & 1.292 & 0.135 & 1.300 & 0.148 & 2.358 & 0.234 \\
                                 & PadAug(HT)                     & 1.170 & 0.122 & 1.261  & 0.144 & 2.311 & 0.229 \\
\midrule
\midrule
\multirow{2}{*}{ECAPA-TDNN-512}
                                 & VAD                        & 1.431 & 0.169 & 1.577 & 0.178 & 2.926 & 0.285 \\
                                 & PadAug(HT)                     & 1.435 & 0.162 & 1.569 & 0.174 & 2.831 & 0.278 \\
\midrule
\midrule
\multirow{2}{*}{ECAPA-TDNN-1024} & VAD                        & 1.223 & 0.142 & 1.345 & 0.155 & 2.515 & 0.249 \\
                                 & PadAug(HT)                     & 1.154 & 0.144 & 1.327 & 0.152 & 2.411 & 0.241 \\
\midrule
\midrule
\multirow{2}{*}{CAM++}           & VAD                        & 1.026 & 0.149 & 1.219 & 0.151 & 2.444 & 0.250 \\
                                 & PadAug(HT)                     & 1.000 & 0.150 & 1.172 & 0.138 & 2.281 & 0.232 \\
\toprule
\end{tabular}
}
\end{table}


\subsection{Evaluation criterion}
In the test stage, we adopted cosine similarity as the scoring function without any score normalization. We employed the minimum detection cost funtion (minDCF) with $P_{target}=0.01$ and $ C_{miss} = C_{fa} = 1 $ and the standard equal error rate (EER) as the evaluation protocols.

\subsection{Main results}
Table \ref{tlb:1} lists the main experimental results of the proposed \textit{PadAug} and baseline on four state-of-the-art SV models. From the table, we see that the proposed \textit{PadAug} improve the system's performance in all cases. For example, for the `Chunk3s' test set of Voxceleb1-O, the ResNet34 model with \textit{PadAug} achieves a relative EER reduction of 5.0\% over that without \textit{PadAug}. Similarly, the relative EER reduction with ECAPA-TDNN-512, ECAPA-TDNN-1024, and CAM++  are 2.4\%, 8.7\%, and 2.2\% respectively.

We also can see that, when the silence period increases, the performance of the SV systems gets worse rapidly, and \textit{PadAug} can effectively reduce the performance degradation. Taking the experimental results on the $'Chunk3s+Head1s+Tail1s'$ test set of Voxceleb1-O as an example. The ResNet34 model with \textit{PadAug} achieves a relative EER reduction of 17.0\% over that without \textit{PadAug}. Similarly, the relative EER reduction with ECAPA-TDNN-512, ECAPA-TDNN-1024, and CAM++ are 19.4\%, 21.7\%, and 19.9\% respectively.

Table \ref{tlb:2} lists the results of the ResNet34 model with the two variants of \textit{PadAug}, i.e. \textit{PadAug(HT)} and \textit{PadAug(HMT)}. From the table, we see that the two variants perform quite similarly, and both outperforms the method without \textit{PadAug}. Note that the performance on the other three representative SV models is similar with that on ResNet34. Due to the length limitation, we omit the result here.

\subsection{Robustness to the length of silence period}

To demonstrate that \textit{PadAug} the advantage of SV is insensitive to the length of silence period in the test data, we controlled the silence-to-speech ratio in a set of $[0/3,1/3,\ldots,8/3]$, where the ratio of e.g. '2/3' denotes that a 3-second chunk was padded with a 2-second silence period.

\begin{figure}[!t]
    \centering
    \includegraphics[scale=0.5]{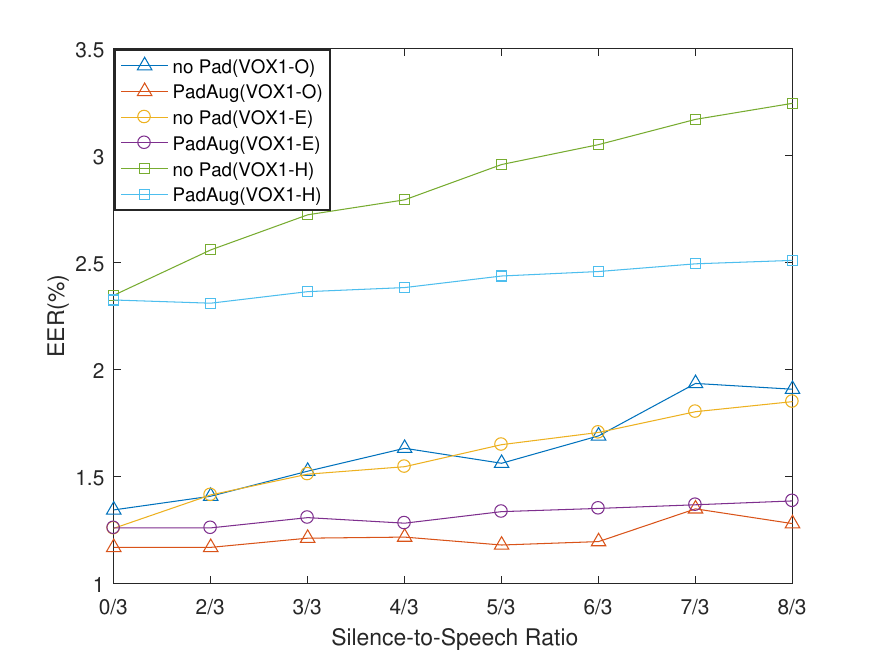}
    \caption{EER (\%) of the proposed method with respect to silence-to-speech ratio. }
    \label{fig:eer}
\end{figure}

Figure \ref{fig:eer} shows the results of the proposed method with respect to the silence-to-speech ratio. From the figure, we can see that, when the silence-to-speech ratio increases, the EER curves of the \textit{PadAug}-based systems are stable, while the EER of the systems without \textit{PadAug} increases rapidly.

\subsection{Comparison to VAD}

In order to show that the proposed \textit{PadAug} overcomes the shortcomings of the VAD-based method, we compared the SV systems that used VAD instead of the proposed \textit{PadAug} with the SV systems that used \textit{PadAug} without VAD. We implemented VAD by webrtcVAD\footnote{https://github.com/wiseman/py-webrtcvad}, which is a widely-used Gaussian mixture model based VAD.

Table \ref{tlb:3} lists the results of comparison on the 'Chunk3s+Head1s+Tail1s' test sets. From the table, we see that the proposed method outperforms or at least does not perform worse than the VAD-based methods. Taking the results on ResNet34 as an example, the proposed \textit{PadAug} outperforms VAD with a relative EER reduction of 9.4\% on the Voxceleb1-O dataset.

\subsection{Comparison to attentive pooling}
From the aforementioned experiments, we see that the proposed method improve the performance of the attentive pooling based SV systems. However, it is known that the attentive pooling itself behaves like an implicit VAD in SV. To compare the proposed \textit{PadAug} with the attentive pooling directly, we firstly used \textit{temporal statistics pooling (TSP)}\cite{snyder2018x} as the pooling layer of the ResNet34 SV model, where the implicit VAD function has been removed. Then, we applied the proposed \textit{PadAug} to the TSP-based ResNet34 SV model.
 
Table \ref{tlb:4} lists the comparison result. From the result, we see clearly that the proposed \textit{PadAug} plus TSP clearly outperforms the attentive statistics pooling (ASP) without \textit{PadAug} in all cases, which indicates that \textit{PadAug} is better than ASP in dealing with short silence periods.

\begin{table}[t]
\caption{Comparison between  attentive statistics pooling (ASP) and the temporal statistics pooling (TSP) with PadAug(HT), where the SV model is ResNet34.  `C+H+T' is short for `Chunk3s+Head1s+Tail1s'. `C+H+T+M' is short for `Chunk3s+Head1s+Tail1s+Mid1s'. }
\centering
\label{tlb:4}
\scalebox{1.0}{
\begin{tabular}{lccccccc}
\toprule
\multirow{2}{*}{Method}  & \multirow{2}{*}{Test set}      & \multicolumn{2}{c}{Voxceleb1-O} & \multicolumn{2}{c}{Voxceleb1-E} & \multicolumn{2}{c}{Voxceleb1-H} \\
\cmidrule(lr){3-4}\cmidrule(lr){5-6}\cmidrule(lr){7-8}
                         &                             & EER            & minDCF         & EER            & minDCF         & EER            & minDCF         \\
\midrule
ASP (no PadAug)             & C+H+T        & 1.409          & 0.142          & 1.415          & 0.156          & 2.560          & 0.249          \\
TSP+PadAug(HT)              & C+H+T       & 1.213          & 0.127          & 1.322          & 0.151          & 2.460          & 0.235          \\
\midrule
\midrule
ASP (no PadAug)               & C+H+T+M & 1.526          & 0.169          & 1.512          & 0.170          & 2.724          & 0.265   \\
TSP+PadAug(HT)              & C+H+T+M & 1.287          & 0.143          & 1.344          & 0.158          & 2.498          & 0.244          \\
\toprule
\end{tabular}
}
\end{table}

\section{Conclusions}
This paper proposed \textit{PadAug} a novel wave-level data augmentation method, to handle the performance degradation of SV system caused by short silence segments that are not easily removable by e.g. VAD. It can be easily described as that the training data is concatenated with short silence segments in the wave level directly. Extensive experimental results on the Voxceleb dataset demonstrate the effectiveness of the proposed method.

\bibliographystyle{splncs04}
\bibliography{mybibliography}
%






\end{document}